\documentclass[sigconf]{acmart}
\AtBeginDocument{%
  }

\setcopyright{acmlicensed}
\copyrightyear{2025}
\acmYear{2025}
\acmDOI{XXXXXXX.XXXXXXX}
\acmConference[AI4DE@KDD'25]{1st Workshop on AI for Data Editing at KDD'25}{August 3, 2025}{Toronto, ON, Canada}

\acmDOI{XXXXXXX.XXXXXXX}
\acmISBN{978-1-4503-XXXX-X/2018/06}

\usepackage[utf8]{inputenc} 
\usepackage[T1]{fontenc}    
\usepackage{setspace}
\usepackage{amsfonts}       
\usepackage{nicefrac}       
\usepackage{xcolor}         
\usepackage{amsmath}
\usepackage{caption}
\usepackage{multirow}
\usepackage{makecell}
\usepackage{subcaption}
\usepackage{color}
\usepackage{xspace}
\usepackage{amsmath}
\usepackage{adjustbox}
\usepackage{varwidth}
\usepackage{booktabs}
\usepackage{soul}
\usepackage{float}
\usepackage{textcomp}
\usepackage{enumitem}
\usepackage{algorithm}
\usepackage{algpseudocode}
\usepackage{tabularray}
\usepackage{wrapfig}
\usepackage{makecell}

\begin{document}

\title{Automated Label Placement on Maps via Large Language Models}

\author{Harry Shomer}
\affiliation{%
  \institution{University of Texas at Arlington}
  \city{Arlington}
  \country{Texas}}
\email{harry.shomer@uta.edu}
\authornote{Research done while an intern at HRL Laboratories}

\author{Jiejun Xu}
\affiliation{%
  \institution{HRL Laboratories}
  \city{Malibu}
  \country{California}}
\email{jxu@hrl.edu}

\renewcommand{\shortauthors}{Shomer et al.}
\renewcommand{\shorttitle}{Automated Label Placement on Maps via LLMs}

\begin{abstract}
Label placement is a critical aspect of map design, serving as a form of spatial annotation that directly impacts clarity and interpretability. Despite its importance, label placement remains largely manual and difficult to scale, as existing automated systems struggle to integrate cartographic conventions, adapt to context, or interpret labeling instructions. In this work, we introduce a new paradigm for automatic label placement (ALP) that formulates the task as a data editing problem and leverages large language models (LLMs) for context-aware spatial annotation. To support this direction, we curate MAPLE, the first known benchmarking dataset for evaluating ALP on real-world maps, encompassing diverse landmark types and label placement annotations from open-source data. Our method retrieves labeling guidelines relevant to each landmark type leveraging retrieval-augmented generation (RAG), integrates them into prompts, and employs instruction-tuned LLMs to generate ideal label coordinates. We evaluate four open-source LLMs on MAPLE, analyzing both overall performance and generalization across different types of landmarks. 
This includes both zero-shot and instruction-tuned performance. 
Our results demonstrate that LLMs, when guided by structured prompts and domain-specific retrieval, can learn to perform accurate spatial edits, aligning the generated outputs with expert cartographic standards. Overall, our work presents a scalable framework for AI-assisted map finishing and demonstrates the potential of foundation models in structured data editing tasks. 
\end{abstract}

\ccsdesc[500]{Computing methodologies~Machine learning}

\keywords{Automated Label Placement, Map Finishing, Large Language Model}

\maketitle

\section{Introduction}

Maps are an omnipresent and vital part of our everyday lives. They provide an intuitive interface for understanding the world around us, allowing us to plan, navigate, and coordinate our future movements. Due to their historical and practical significance, the discipline of cartography has emerged as a way of formalizing and studying the mapmaking process. The primary goal of cartography is to produce maps that are both accurate and easy to interpret.

\begin{figure}
    \centering
    \includegraphics[width=1\linewidth]{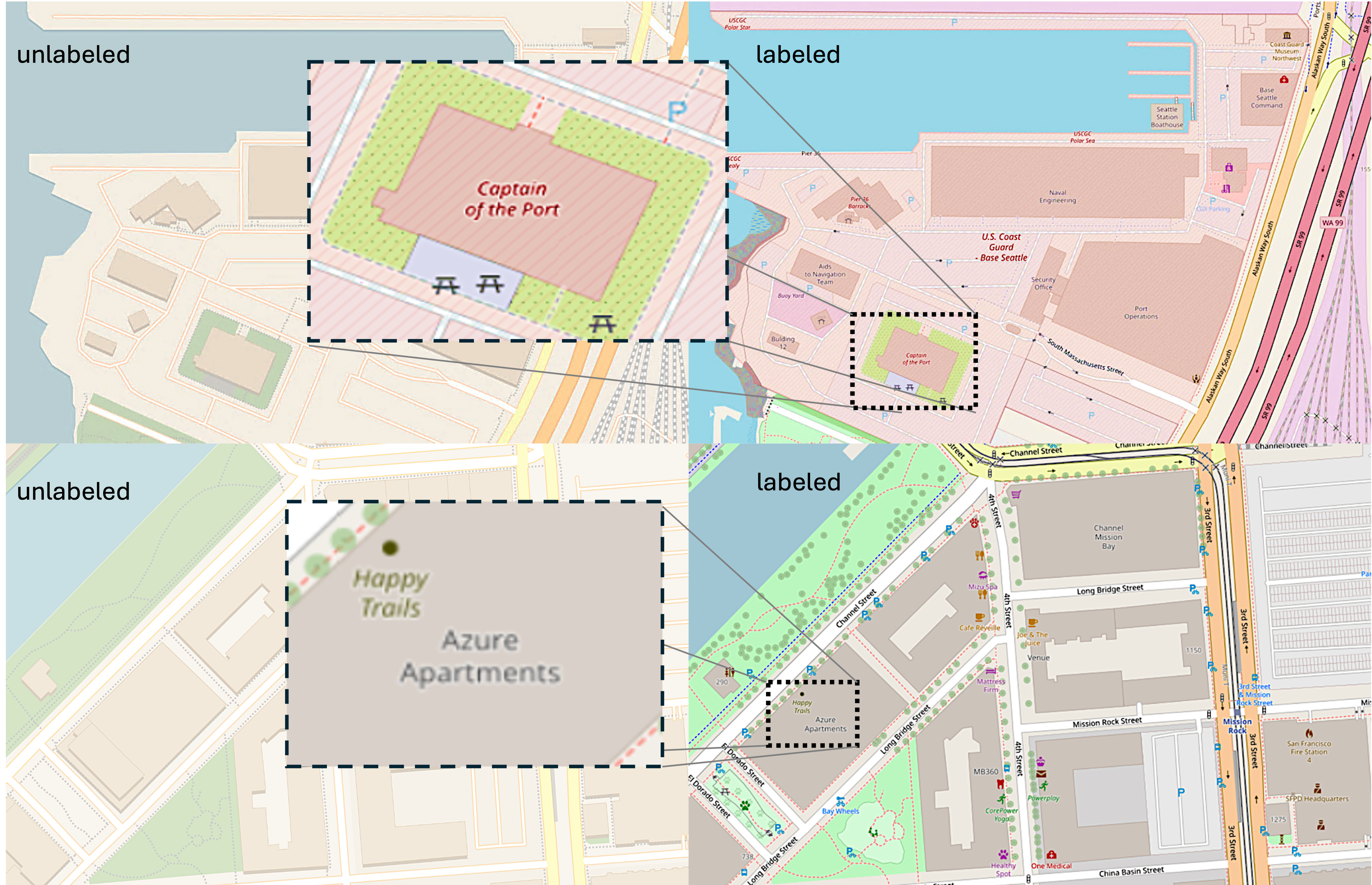}
    \caption{Illustration of label placement on maps. The two left subfigures show unlabeled maps, while the two right subfigures show the corresponding labeled versions. The zoomed-in insets highlight examples of labels positioned near landmarks. The goal of our work is to develop an automated and scalable approach for placing labels at appropriate locations on the map, respecting labeling guidelines, and adapting to the surrounding context.}
    \label{fig:label_placement_example}
\end{figure}

One central problem in cartography is label placement \cite{psu_maps_geospatial,field2018cartography}. 
Figure~\ref{fig:label_placement_example} illustrates the concept of label placement by comparing unlabeled maps (left) with their labeled counterparts (right). Positioned near each landmark is a label placed in a contextually appropriate manner.
Placing these labels is a subtle but essential aspect of map design. It requires balancing spatial proximity to the referenced feature with overall readability and visual clarity. To ensure consistency and legibility, cartographers typically follow detailed guidelines that prescribe label positioning based on the type of landmark and surrounding context.
However, manual label placement is often tedious and time-consuming, as it requires examining each feature individually and applying context-specific rules.
This challenge has also been recognized at the national scale: the National Geospatial-Intelligence Agency (NGA), responsible for producing mission-critical maps, recently issued a $\$708$ million contract for large-scale geospatial data labeling to support automation in map production \cite{nga_announcement}. This investment underscores the complexity and scale of the labeling problem in modern cartographic workflows, particularly for mission-critical geospatial applications.

To address these challenges, {\bf automatic label placement} (ALP) \cite{bobak2023reinforced, yu2024deep, oucheikh2024feasibility, niedermann2017automatic} has emerged as a computational method for determining label positions algorithmically. 
Many solutions exist that consider the problem of ALP on maps,
including rule-based engines embedded in Geographic Information Systems (GIS) software, optimization-based formulations (e.g., integer programming or heuristic search), and more recent machine learning methods such as deep reinforcement learning, generative models, and graph-based neural networks. These systems aim to maximize label coverage while minimizing visual conflicts such as overlaps and misalignment. Some approaches optimize label positions based on geometric criteria, while others attempt to learn placement patterns from annotated datasets.
However, these methods tend to be limited in several respects. {\bf First}, they are often inflexible, requiring significant manual configuration or retraining to adapt to new map requirements or domain-specific conventions. {\bf Second}, they struggle to simultaneously handle the many constraints found in real-world labeling scenarios, especially on dense or heterogeneous maps. {\bf Third}, and perhaps most importantly, they lack the ability to interpret and apply textual labeling guidelines, such as those found in cartographic guidebooks, that describe nuanced preferences for different landmark types, contextual relationships, or visual hierarchies. As a result, these systems fall short in producing label placements that are both scalable and semantically aligned with human-designed cartographic standards.

These observations motivate us to ask -- {\it Can we design a method for ALP that can efficiently consider a set of human-readable guidelines in its reasoning process?}. To tackle this problem, we attempt to use Large Language Models (LLMs)~\cite{grattafiori2024llama, team2024gemma}, to determine the label placement. We consider the use of LLMs due to their ability to flexibly reason on a variety of situations in a relatively efficient manner. To incorporate the relevant labeling instructions from a set of human-readable guidelines, we consider the use of Retrieval Augmented Generation (RAG)~\cite{rag}. Our RAG pipeline is designed to retrieve the relevant instructions for a given landmark to aid the LLM in placing the label. To better orient the LLMs to the task of ALP, we also try instruction tuning LLMs for the task of map labeling via LoRA~\cite{lora}. We also observe that there {\bf is no open-source dataset benchmarking method for automatic label placement on maps}. To remedy this issue, we design a new dataset for this problem, {\bf MAPLE} – {\bf M}ap {\bf A}utomatic {\bf PL}acement {\bf E}valuation. MAPLE contains 100 maps from three major cities and over 1000 individual landmarks that require labeling. The data is extracted from the popular OpenStreetMap~\cite{osm}. We further consider the use of an external set of labeling guidelines to guide the label placement on these maps.
Our contributions can be summarized as the following:
\begin{enumerate}
    \item We create the first open-source dataset for benchmarking automatic label placement on maps -- {\bf MAPLE}. 
    \item We design new strategies for both retrieving relevant guidelines for a landmark using RAG and instruction tuning LLMs for ALP. 
    \item Through extensive experiments, we benchmark the ability of four different open-source LLMs on our new dataset. We further test their ability by type of landmark and various prompt designs. 
\end{enumerate}
\section{Background and Related Work} \label{sec:related}

Automatic label placement is a longstanding challenge in cartography and information visualization, involving the positioning of textual annotations on maps and diagrams to maximize readability and minimize overlap. Traditional methods have evolved over time, incorporating rule-based heuristics, optimization techniques, and, more recently, machine learning approaches. We categorize existing work into the following four primary areas.

\textbf{Rule-based Approaches}: 
Rule-based approaches have historically served as the cornerstone of automated label placement in cartography and geographic information systems (GIS). These systems formalize long-standing cartographic design principles, such as minimizing label overlaps, maximizing proximity to labeled features, and preserving legibility, into a sequence of deterministic heuristics. One of the most influential early frameworks for these rules is based on the cartographic theory developed by \citet{imhof1975positioning}, who outlined qualitative principles for label prioritization, typeface choice, and positioning relative to points, lines, and areas.
Contemporary GIS software packages, such as Esri’s ArcGIS Pro, implement these principles through the Maplex Label Engine, a proprietary module that offers advanced controls for conflict resolution, label placement strategy (e.g., curved versus horizontal text), and feature-specific behavior (e.g., roads versus rivers) \cite{esri_maplex}. Similarly, the open-source QGIS project includes the PAL labeling engine, which provides a flexible rule-based system with configurable priorities and spatial constraints  \cite{qgis_pal}. 
Major web‑mapping platforms like Google Maps employ sophisticated rule‑based engines to prioritize feature types, detect collisions, and index spatial data to automatically place labels during panning and zooming operations \cite{google_maps_jsapi}.
These engines enable dynamic label positioning that adapts to map scale and content density and are widely adopted in professional cartographic workflows. While rule-based systems are highly effective for standard mapping tasks, they inherently lack semantic understanding of the features they label. They operate on geometric constraints, without incorporating contextual knowledge about landmark importance, category, or inter-feature relationships. Our approach fundamentally differs from traditional rule-based systems by leveraging large language models (LLMs) to interpret and apply labeling instructions dynamically. This enables greater adaptability and semantic awareness, allowing for more nuanced label placements that are sensitive to both cartographic conventions and the specific contextual properties of a given map.

\textbf{Optimization and Mathematical Programming}: 
Beyond rule-based systems, optimization-based approaches offer a more global perspective on the label placement problem by formulating it as a constrained optimization task. These methods aim to discover optimal or near-optimal label configurations that satisfy multiple constraints, such as non-overlap and proximity to anchor features, while maximizing objectives like label coverage or legibility. A wide range of classical optimization paradigms have been explored in this space, including Integer Linear Programming (ILP), Quadratic Programming, Simulated Annealing, and Genetic Algorithms \cite{christensen1995empirical}. These methods treat each label as a variable whose placement is subject to constraints, and they search for a configuration that minimizes a global cost function or maximizes layout utility. For instance, metaheuristic techniques like the Hybrid Discrete Artificial Bee Colony (HDABC) algorithm have shown promise in navigating the vast search space of potential label positions efficiently \cite{cao2023hybrid}.
A particularly comprehensive treatment of these approaches is provided \citet{niedermann2017automatic}, which systematically studies the use of exact and approximate optimization techniques for various types of map features, including point, line, and area labeling. Their work introduces a modular framework that supports different mathematical formulations, including Mixed-Integer Programming and Constraint Satisfaction, and presents empirical evaluations across multiple real-world datasets. Notably, they highlight the trade-offs between optimality and computational feasibility, showing that while exact methods often yield high-quality results, they can be impractical for dense or large-scale maps without specialized heuristics or problem decomposition.
While these optimization techniques offer significant improvements over purely rule-based systems, they remain fundamentally geometric in nature. They typically do not incorporate semantic, contextual, or user-driven considerations into the placement process.
Our work departs from these methods by leveraging LLMs to synthesize placement instructions based on both spatial attributes and semantic metadata, enabling a more adaptable, context-aware approach that goes beyond geometric optimization.

\textbf{Deep Learning for Visual Label Prediction}
Recent advances in deep learning have opened new directions for automating label and layout generation tasks by learning spatial patterns directly from data. In contrast to rule-based or optimization approaches, deep learning methods model layout prediction as a supervised or reinforcement learning task, using neural networks to learn spatial relationships, feature importance, and layout aesthetics from labeled examples.
One early direction involves using convolutional neural networks (CNNs) and generative models to synthesize label positions. For example, \citet{oucheikh2024feasibility} explored the use of GANs (Generative Adversarial Networks)~\cite{goodfellow2014generative} trained on expert-labeled maps to predict plausible label layouts. They show promising results in reproducing stylistic elements of cartographic design. Another line of work models the label placement problem as a graph reasoning task. For example, \citet{qu2024graph} proposed a Graph Transformer architecture that treats labels and landmarks as nodes and learns to place them based on their spatial and semantic relationships.
Reinforcement learning has also been applied to label placement through a learning-based optimization lens. In particular, \citet{bobak2023reinforced} introduced a multi-agent deep reinforcement learning (MADRL) framework where each label acts as an autonomous agent. These agents learn through trial-and-error to maximize overall label completeness and minimize collisions, outperforming conventional rule-based and heuristic strategies in dense feature maps.
Beyond cartographic applications, recent work has demonstrated the potential of large language models (LLMs) for general-purpose visual layout generation. LayoutGPT \cite{feng2023layoutgpt} treats layout generation as a visual planning task, using in-context learning with LLMs to produce complex web-style layouts from textual prompts. Similarly, Design2Code \cite{si2024design2code} benchmarks the ability of multimodal LLMs such as GPT-4V and Gemini Vision Pro to convert visual designs into HTML/CSS code, showing that LLMs can reason about layout structure and spatial alignment with minimal task-specific supervision.
Our approach builds on this line of work by introducing a retrieval-augmented prompting method that incorporates landmark-specific labeling guidelines into a language model to predict ideal label coordinates. By combining spatial inputs (e.g., visual coordinates) with semantic metadata (e.g., landmark types), our method extends deep learning beyond layout reproduction toward context-aware, convention-compliant label planning in map finishing.

\textbf{Cognitive Studies on Label Preferences}
In addition to algorithmic correctness, effective label placement also depends on how humans perceive and interpret spatial arrangements. A growing body of research in cartographic design and visual cognition highlights that factors such as alignment, spacing, and grouping significantly influence the usability and readability of labeled maps.
For example, Scheuerman et al. \cite{scheuerman2023visual} conducted user studies showing that participants consistently preferred label placements with clearer alignment and visual symmetry, even when these conflicted with purely geometric optima. 
Similarly, Bobák et al. \cite{bobak2024top} found that users overwhelmingly favored labels placed directly above point features, contradicting traditional top-right placement conventions long assumed to be optimal.
Such findings suggest that traditional evaluation metrics like overlap minimization or proximity alone are insufficient for ensuring perceptual quality.

These insights expose a key limitation of many rule-based and optimization-driven systems: while they may produce technically valid layouts, they often fail to account for human-centered criteria such as aesthetic balance or visual hierarchy. Efforts to incorporate user feedback or style preferences into labeling systems exist but are often domain-specific and require manual configuration or retraining.
Although our method does not explicitly model human preferences, this line of work underscores the need for more adaptive, semantically informed approaches to layout generation. By moving beyond rigid rules or fixed optimization objectives, we aim to create a system that can more flexibly respond to diverse labeling goals, potentially including human-centric instructions in future iterations. The growing body of perceptual research provides a valuable foundation for such extensions and reinforces the need to consider not just what is correct, but also what is visually effective.

\section{Dataset Construction} \label{sec:new_datasets}

In this section we detail the construction of a new dataset for benchmarking the task of automatic label placement on maps, {\bf MAPLE} -- {\bf M}ap {\bf A}utomatic {\bf PL}acement {\bf E}valuation. The dataset is constructed from OpenStreetMap (OSM)~\cite{osm} and contains 100 maps from multiple cities. Each map contains, on average, 13 different landmarks that require labeling, with a total of 1276 total landmarks. An example of two maps is shown in Figure~\ref{fig:label_placement_example}. To guide the label placement for each landmark, we consider the use of a publicly available label placement guideline. 

The dataset construction process contains three main steps: {\bf (a)} Collecting the raw data from OSM, {\bf (b)} Determining the label location for each landmark, {\bf (c)} Identifying and processing a set of labeling guidelines. In the rest of this section, we detail each step. 

\begin{figure*}
  \centering
  \includegraphics[width=0.9\linewidth]{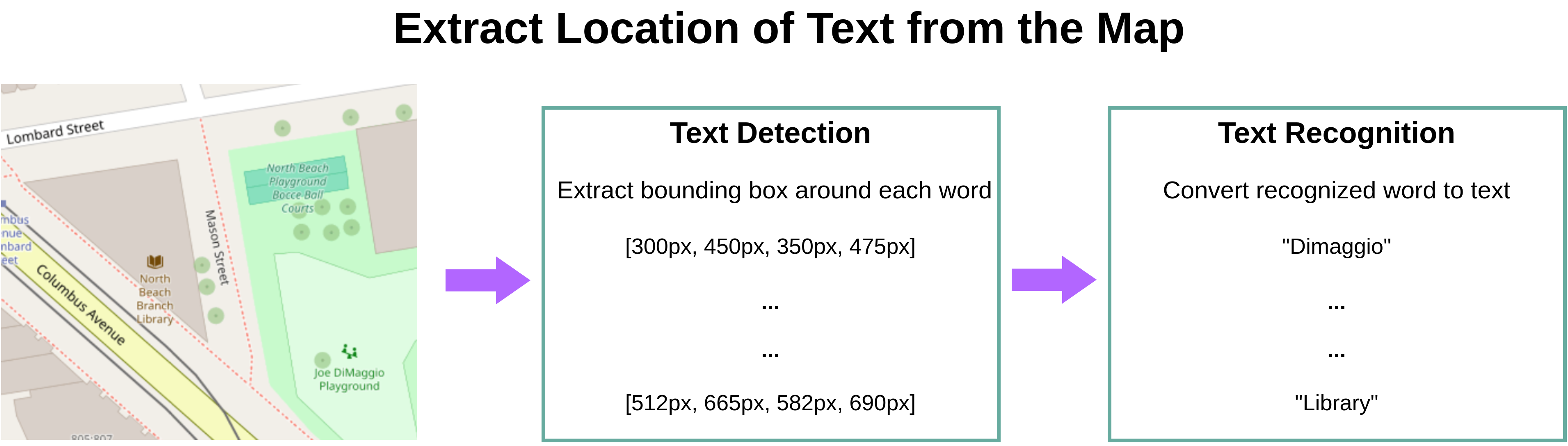}
  \caption{The pipeline for extracting the label text from a map. For a map we first detect the location of all text in the map. Then for each piece of recognized text, we perform optical character recognition (OCR) to convert it to a machine-readable format.}
  \label{fig:lbl_pipeline1}
\end{figure*}

\subsection{Data Collection using OpenStreetMap} \label{sec:osm_data}

The first step to building our new dataset MAPLE is to collect a set of maps. We consider a map to be a small area that contains roughly 10-20 different landmarks. For each map, we require the following pieces of information to test any automatic label placement method:
\begin{enumerate}
    \item An image of the map with the landmark labels included.
    \item An image of the map {\bf without} the labels included. This allows us to generate new maps with different algorithms.
    \item The name and location of each landmark in the map. The location is expressed as a set of $k$ coordinates, which is necessary to account for the variety of different shapes a landmark can take.
    \item The location of the corresponding text for each landmark on the map (i.e., ``the label''). 
    \item The type of landmark (e.g., an ``office'' or ``shop''). This is necessary as labeling instructions often differ based on landmark type.
\end{enumerate}

For retrieving this information, we use OpenStreetMap (OSM)~\cite{osm}, which is a free and publicly available platform that provides accurate map data, including detailed information about landmarks in a given region.

Once we identify a single region that we wish to extract, we perform the following steps. First, we use QGIS~\cite{qgis_pal} to retrieve an image of the map {\it with} and {\it without} the labels of each landmark. QGIS is a free geographic information software that is used for viewing or manipulating various types of map data. We use QGIS for this step as it is highly customizable, allowing us to retrieve images of maps without the labels. This is not possible with the native OSM software. 
Second, we query the OSM API~\cite{osm} using the region's coordinates to extract metadata for each landmark. This includes their name, geographic location, and type.
By default, OSM uses the EPSG coordinate system to identify each landmark and map. However, since each map is only a small region, the use of a global coordinate system is unnecessary. As such, we replace the coordinates of each landmark with its pixel values for that map's image. This allows us a much simpler and easier identification system for determining the location of landmarks for a single map. Finally, to focus the dataset and reduce complexity, we restrict our extraction to a curated set of commonly occurring landmark types. In total, we include 7  landmark categories, as summarized in Table~\ref{tab:dataset_stats}, which also reports the number of instances collected for each type.

\begin{table}[h]
\centering
 \caption{Dataset Statistics by Landmark Type. 
 }
 \vskip -1em
    \begin{tabular}{c|c}
    \toprule
     {\bf Type} & {\bf \# Landmarks} \\
     \midrule
    {\bf Tourism} & 176 \\
    {\bf Shop} & 209 \\
    {\bf Amenity} & 572 \\
    {\bf Leisure} & 94 \\
    {\bf Office} & 62 \\
    {\bf Building} & 491 \\
    {\bf Place} & 15 \\
    \bottomrule
    \end{tabular}
    \label{tab:dataset_stats}
\end{table}

The above procedure is repeated 100 times for different areas.
To enhance the diversity of the maps we extract, we consider maps from three different locations. This includes Seattle, Los Angeles, and San Francisco. The number of maps from each city, along with the number of total landmarks, are shown in Table~\ref{tab:dataset_loc}. 

\begin{table}[h]
\centering
 \caption{Dataset Statistics by City. 
 }
 \vskip -1em
    \begin{tabular}{c|cc}
    \toprule
     {\bf City} & {\bf \# Maps} & {\bf \# Landmarks} \\
     \midrule
    {\bf Los Angeles} & 34 & 605 \\
    {\bf San Francisco} & 33 & 562\\
    {\bf Seattle} & 33 & 453\\
    \bottomrule
    \end{tabular}
    \label{tab:dataset_loc}
\end{table}

\begin{figure*}
  \centering
  \includegraphics[width=0.7\linewidth]{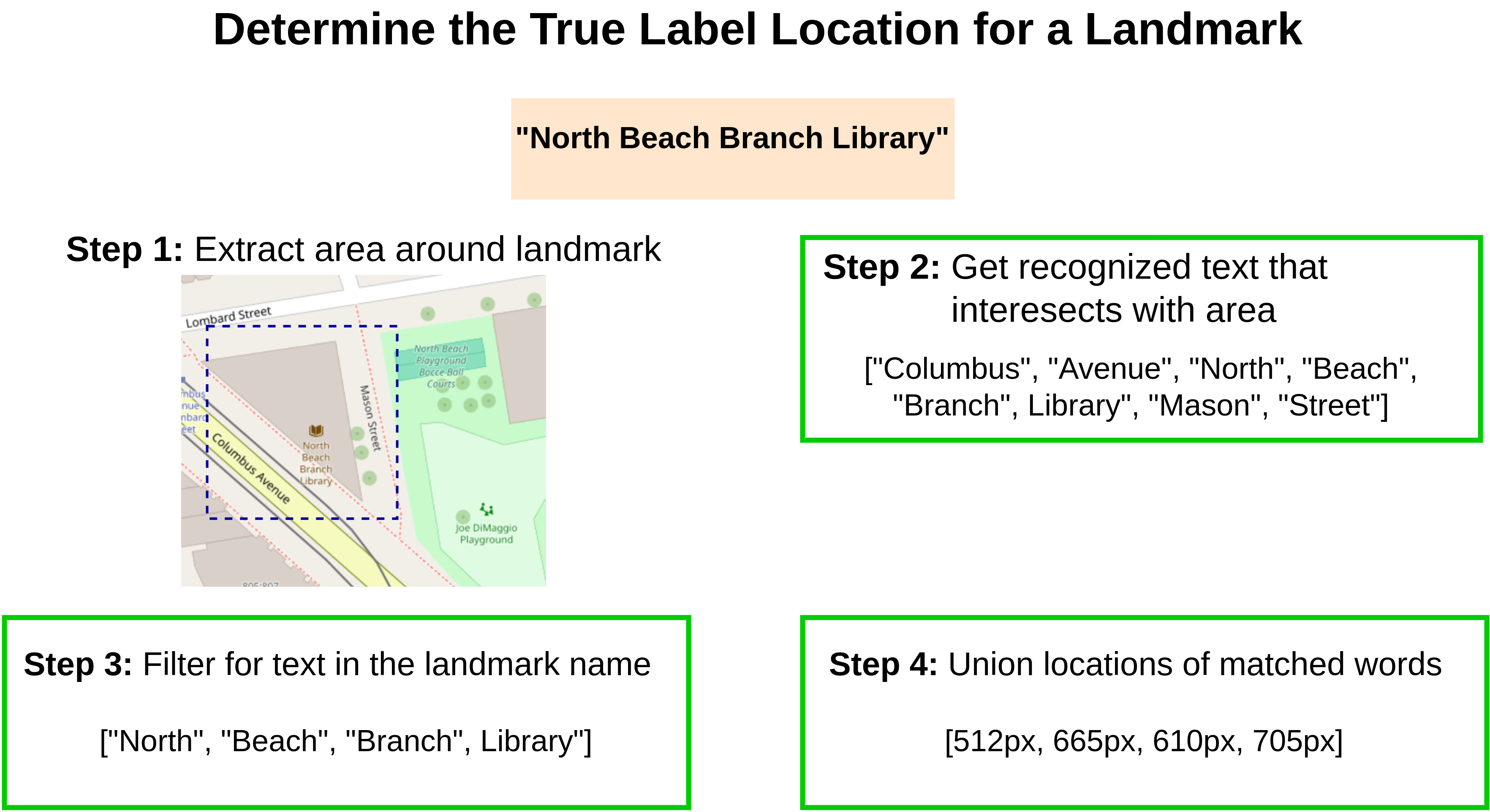}
  \caption{The pipeline for determining the true label location for a single landmark. We first (1) extract the area around the landmark in question (2) find all recognized text (shown in Figure~\ref{fig:lbl_pipeline1}) in that area (3) filter for text found in the landmark's name (4) union the location of all identified text to form the final label location.}
  \label{fig:lbl_pipeline2}
\end{figure*}

\subsection{Determining the Correct Label Location}

In the Section~\ref{sec:osm_data} we discussed how the raw data is retrieved from OSM. However, one piece of information that we can not retrieve via OSM is the location of each landmark's label on the map. This is because the labeling system used by QGIS (i.e., PAL)~\cite{qgis_pal} is not exposed via their API. For example, in Figure~\ref{fig:label_placement_example} we can see that the map contains the label ``Azure Apartments'' to mark the location of that landmark. The precise location on the map of where the label ``Azure Apartment'' is vital, as it serves as the ground truth for any ALP method. However this location is not accessible from OSM. 
As such, in order to evaluate any ALP method, {\it we must determine a strategy for determining the ground truth location of each landmark's label}.

To overcome this problem, we consider a text detection and recognition pipeline to appropriately determine the label for each landmark. Specifically, given a single map, we perform the following steps: {\bf (a)} Detect all text in the map, {\bf (b)} For each piece of detected text, we use Optical Character Recognition (OCR) to convert it to a machine-readable format, {\bf (c)} Assign each piece of recognized text to the proper landmark, {\bf (d)} Determine the ground truth label location for a landmark as the union of all detected text assigned to it. We now detail each step in this pipeline. 
\newline\newline
\noindent{\bf Text Detection}: We first must determine the location of all the text in a given map. To achieve this, we use DBNet++~\cite{liao2020real}. DBNet++ is a text detector which proposes a  differentiable binarization module to enhance text detection capabilities. The output of this algorithm is a bounding box that describes the location of any piece of text identified in the map.
\newline\newline
\noindent{\bf Text Recognition}: Now that we've identified {\it where} all the text on a map is located, we must now determine {\it what} that text says. Currently, the text is simply part of an image, which while readable for a human, is not by a machine. To recognize the text we use  ABINet~\cite{fang2021read}. ABINet uses a vision transformer along with a bidirectional language model to recognize a piece of text. We run ABINet on each piece of detected text for a map. An example of the text detection and recognition pipeline for a single map is shown in Figure~\ref{fig:lbl_pipeline1}.
\newline\newline
\noindent{\bf Text Assignment}: We now have the location and actual text of all the words on a map. However, {\it how do we assign each word to the correct landmark}? For example, in Figure~\ref{fig:lbl_pipeline1} we can see that the words ``Columbus Avenue'' are right next to the landmark for the ``North Beach Branch Library''. How do we determine which words are the correct text for the library's label? To achieve this we first assume that the label must be nearby it's landmark. Specifically we assume that it must be no more that $p$ pixels from the boundary of the landmark. This is reasonable as labels almost always either intersect with the landmark itself or are very close by. In practice we find that p=50px works well. Next, we extract all detected text that intersect with the landmark. For example, in Figure~\ref{fig:lbl_pipeline1} for ``North Beach Branch Library'' this might be the words [``Columbus'', ``Avenue'', ``North'', ``Beach'', ``Branch'', ``Library'', ``Mason'', ``Street'']. We note that a single detected word can only be assigned to one landmark. We then check if the intersecting words are contained in the landmark's name. In practice, we use fuzzy matching instead of exact matching, as the output of any OCR method may contain some errors. We specifically use Levenshtein distance with an 80\% distance threshold. We give an example of this process in Figure~\ref{fig:lbl_pipeline2} in steps 1-3.
\newline\newline
\noindent{\bf Ground Truth Label Determination}: For every landmark in a map, the previous step gives us a set of nearby words that are contained in the landmark's name. Furthermore we have the location of each word in the image, in the form of a bounding box, via text detection. We can then use that information to determine the ground truth locations of the landmark's label by simply combining the location of each word. Let us assume that a landmark is assigned $n$ words with locations $[w_1, w_2, \cdots, w_n]$. The ground truth location of the label, $l$ is given by the union of each such that $l = \cup_{i=1}^n w_i$.

In practice we are not able to determine the ground truth label location for all landmarks. This is because either the recognized text: (a) wasn't detected, (b) is far away from the landmark, (c) is too different from it's true value (due to poor text recognition). With that said, in practice we find that this pipeline is highly effective in determining the label location for most landmarks. Specifically, we found that for about {\bf 87\% of maps}, we are able to determine a label location, thus validating our overall pipeline.

\subsection{Labeling Guidelines}

\begin{figure*}[h!]
  \centering
  \includegraphics[width=0.95\linewidth]{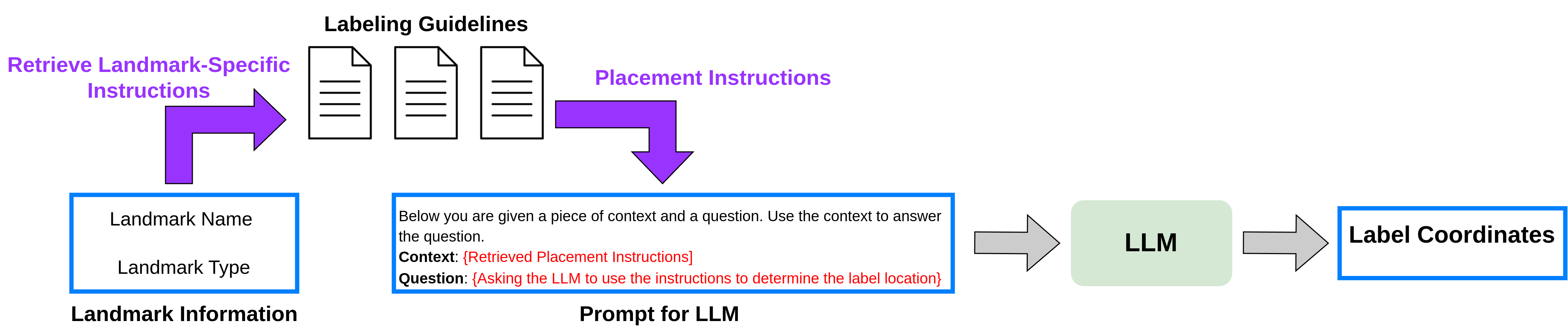}
  \caption{Overall framework for performing ALP using LLM+RAG. We first retrieve the correct instructions using both the name and type of the landmark. This information is then included in a prompt to the LLM, along with the location of the landmark, for the LLM to determine where to place the label.}
  \label{fig:framework}
\end{figure*}

In the last two sections we discussed how we constructed our dataset from OpenStreetMap~\cite{osm}. This includes, the collection of the individual maps, landmark location, and our pipeline for determining the location of each landmark's label. However, an important question to consider is {\it how do we guide any algorithm to properly determine the location of a label}?

As previously discussed in Section~\ref{sec:related} many existing approaches either use rule-based approaches~\cite{qgis_pal}, optimization techniques~\cite{niedermann2017automatic}, or visual label prediction~\cite{feng2023layoutgpt}. However, a drawback of these approaches is that they all rely on hard constraints, and with any change in the underlying label placement guidelines, the approach itself needs to be either modified. As such, we consider a different and more flexible approach to determining label placement. Instead, we consider the use of {\bf human-readable guidelines for label placement}. These guidelines can they be used by a model to guide their reasoning when deciding where to place a label. 
Such guidelines are common in practice, as they are often used by professional organization to standardize their procedure for map labeling.


For the choice of guidelines, we use those published by the National Geospatial-Intelligence Agency (NGA)~\cite{ngaDPS}. These guidelines are used by the United States to create accurate maps for national security operations. Later in Section~\ref{sec:vector_db} we describe how these guidelines are processed and stored by our proposed methods.

\section{Methods} \label{sec:methods}

In the previous section we introduced a new dataset, {\bf MAPLE}, for benchmarking automatic label placement (ALP). To guide the correct label placement, MAPLE includes a set of textual labeling guidelines that describe how different landmarks should be labeled. However, as discussed earlier in Section~\ref{sec:related}, previous approaches to ALP are unable to flexibly use a set of textual guidelines to steer their decision making process. In the following subsection we introduce our proposed solution for overcoming this issue. 

\subsection{Overall Design} \label{sec:basic_design}

To enable the task of ALP given a set of textual guidelines, we consider the use of Large Language Models (LLMs)~\cite{grattafiori2024llama, team2024gemma}. This is due to LLMs strong ability to understand textual information and perform various reasoning tasks. To incorporate the use of the external labeling guidelines, we use LLMs in conjunction with Retrieval Augmented Generation (RAG)~\cite{rag}. Specifically, our RAG pipeline first stores the instructions in a vector database. Then when labeling a specific landmark, we query the database to retrieve the instructions that are personalized to that landmark. The information about the landmark and the retrieved guidelines is then passed to an LLM, which outputs the suggested coordinates of our label. An overview of the framework is given in Figure~\ref{fig:framework}.

In designing this framework, there are a few questions that must be considered. {\bf (1)} How do we process the labeling guidelines for storage in a vector database? {\bf (2)} How do we retrieve the relevant instructions for a specific landmark? {\bf (3)} How do we most optimally prompt the LLM for our task? {\bf (4)} Can we fine-tune the LLMs for better downstream performance? In the following subsections we answer these questions.

\subsection{Vector Database Construction} \label{sec:vector_db}

In order to retrieve the pertinent instructions via RAG, we must have a way of storing the instructions. Vector databases~\cite{vec_dbs} are a common method of doing so. They operate by storing each document as a vector, where the vector is often an encoded version of the underlying data. By storing the data as vectors, we enable quick retrieval through vector similarity measures.

In order to store the guidelines in a vector database, we first must decide at what granularity the vectors should be. A common method in RAG is to chunk the data into fixed chunks, where each chunk contains $C$ tokens. However, a concern with this method is that many instructions may have a length longer or shorter than $C$. As such, using a fixed chunk strategy may result in us splitting certain labeling instructions into multiple chunks or combining multiple instructions into one vector. Therefore, {\it a fixed chunk size strategy is not optimal for our task}.

Instead, we choose to encode each section as its own vector. As such, each entry in our database corresponds to one section of instructions. This ensures that a single vector contains all the labeling instructions of only one type of landmark. Given the text of each section, they are encoded using the nomic-embed-text~\cite{nussbaum2024nomic} text embedder.

\subsection{Prompting LLMs for Labeling} \label{sec:method1}

Using an LLM in conjunction with RAG requires two important considerations. First, we must design a strategy for retrieving the most accurate set of instructions from our vector database for each landmark. Second, careful consideration needs to be given to the prompt design for our task. This is crucial, as the prompt can often have a strong impact on the performance in many tasks.

First, to retrieve the correct instructions, we use both the name and type of the landmark. This information is encoded using the nomic-embed-text text embedder~\cite{nussbaum2024nomic}. To identify relevant instructions, we perform a vector similarity search against our instruction database and return the top $k$ most similar entries. These instructions are then re-ranked based on their relevance to the specific landmark context, considering factors such as keyword overlap, instruction specificity, or optional metadata if available. The text of the top $k$ instructions is then concatenated and appended to the prompt for use by the LLM.

\begin{figure}[h]
  \centering
  \includegraphics[width=0.95\linewidth]{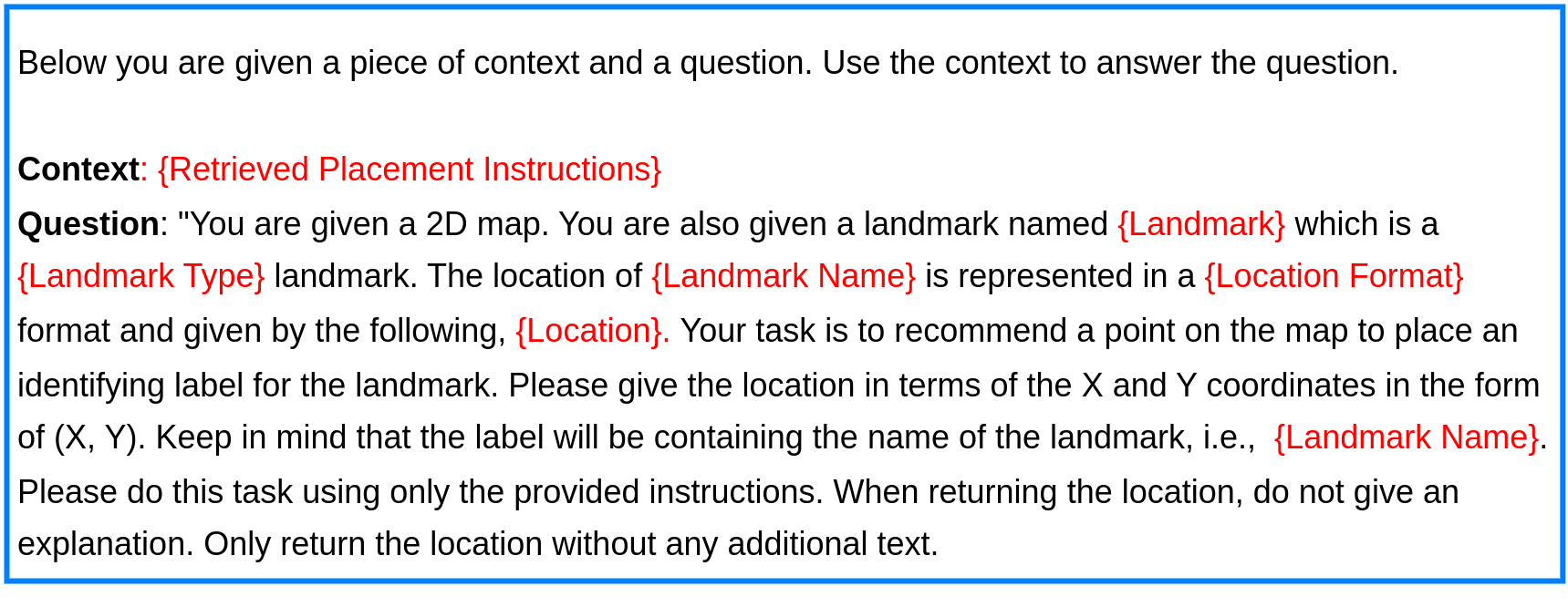}
  \caption{Prompt used to ask LLM for labeling. Those values in \textcolor{red}{red} are specific to each landmark.}
  \label{fig:std_prompt}
\end{figure}

With the corresponding instructions, we are now ready to prompt the LLM to determine the location to place the label. The design of the prompt is shown in Figure~\ref{fig:std_prompt}. We denote the landmark-specific components in red. This includes the retrieved instructions, the name and type of the landmark, the location of the landmark, and how we format the location. As a reminder, the location is composed of a set of coordinates used to mark the boundary of the location and is retrieved from OpenStreetMap. Crucially, we try multiple strategies for formatting the location. This is because a specific LLM may be suited to understand certain formats over others when determining the location on an image. For example, previous work has shown that LLMs can generate CSS for the task of visual planning~\cite{feng2023layoutgpt}. We try four different formatting strategies, where the coordinates are represented as: List, JSON, CSS, XML. For example, in the list format, the coordinates may look like ``$[(100, 150), (250, 300), (100, 400), (250, 500)]$''. Later, in Section~\ref{sec:exp_main}, we find that the type of format can indeed have a large impact on performance.

\subsection{Instruction-tuning LLMs for Labeling} \label{sec:method2}

In the previous section, we describe a method for prompting LLMs to place the label of landmarks given a set of retrieved instructions. However, a concern is that current LLMs may not be suitably optimized for this task. As such, using the existing LLMs may result in poor performance. Due to this, we consider tuning each LLM to better align it with our task.

\begin{figure}[h]
  \centering
  \includegraphics[width=0.95\linewidth]{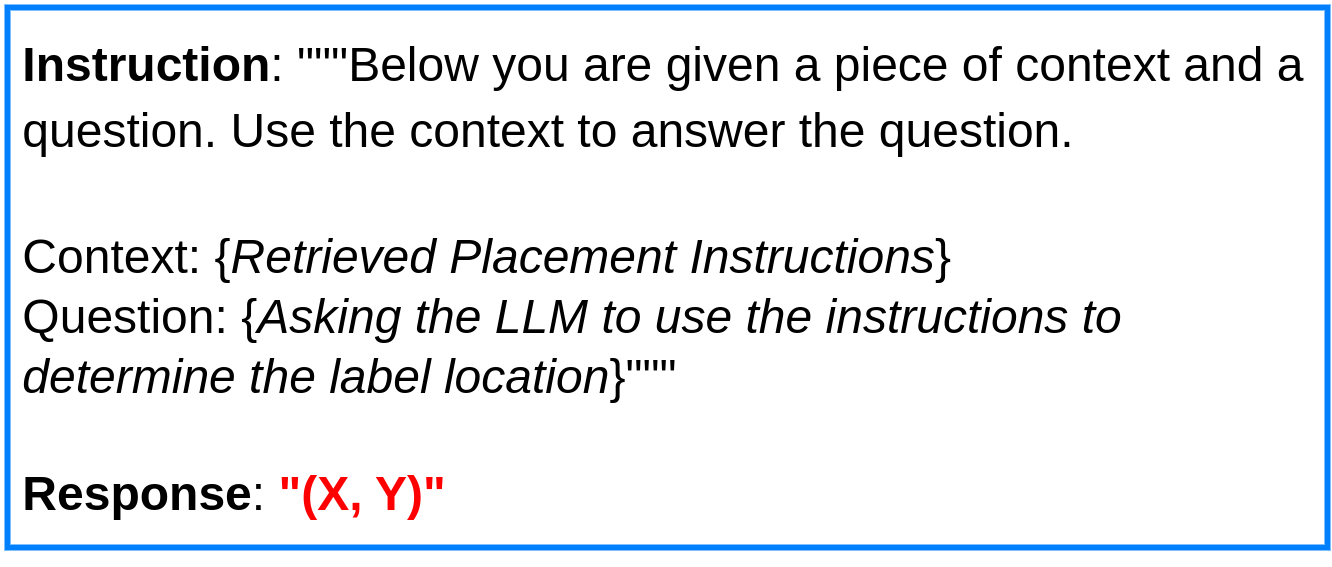}
  \caption{Format of instruction tuning strategy. The LLM is trained to respond to the instruction with the correct (X, Y) coordinates, which correspond to the ground truth location of the label.}
  \label{fig:qlora_prompt}
\end{figure}

We specifically consider instruction tuning each LLM for our task~\cite{zhang2023instruction}. Instruction tuning involves giving an LLM a set of instruction-response pairs, where the LLM takes the instruction as input and must learn to give the appropriate response. In our case, the instruction will be the prompt shown in Figure~\ref{fig:std_prompt}, and the response is the output (X, Y) coordinates of where to place the label. We shown an example in Figure~\ref{fig:qlora_prompt} where the LLM must learn to respond to the {\bf Instruction} with the correct {\bf Response}.

In order to tune each LLM, we use LoRA~\cite{lora}. Instead of modifying the original weights of the LLM, LoRA instead learns an additional set of low rank weight matrices that are then added to the original weights of the LLM. This allows for much more efficient tuning due to only having to learn much fewer weights. We further consider the quantized version of LoRA, QLoRA~\cite{dettmers2023qlora}, to further enhance the training efficiency.  

\begin{figure}[h]
  \centering
  \includegraphics[width=0.95\linewidth]{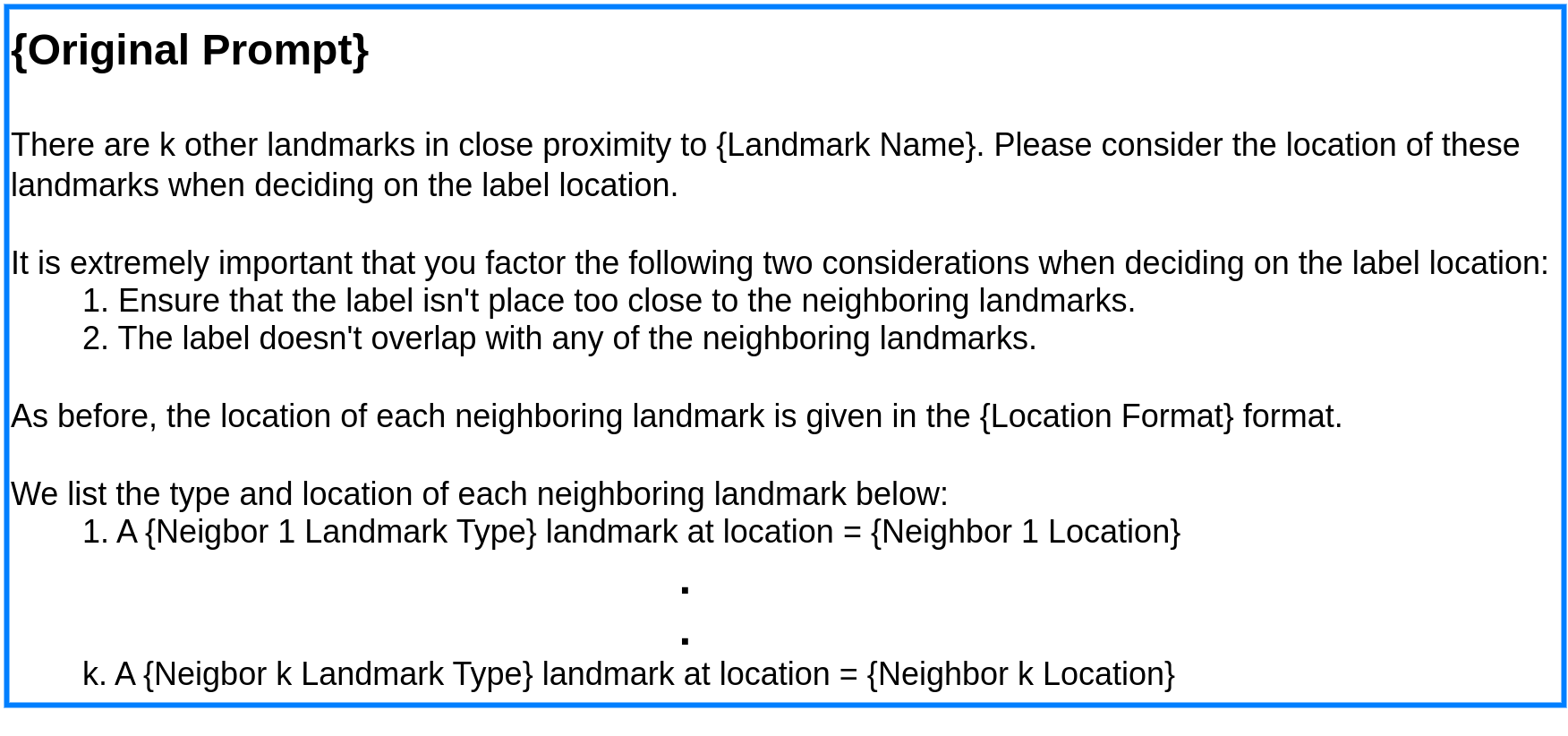}
  \caption{Format of the prompt when including neighboring landmarks. This is appended to the original prompt in Figure~\ref{fig:std_prompt}.}
  \label{fig:neighbor_prompt}
 \vskip -1em
\end{figure}

\subsection{Including Neighboring Context} \label{sec:method_neighbors}

\begin{table*}[h]
\centering
 \caption{Overall Results (RMSE) by LLM, format, and strategy.}
 \vskip -0.5em
\begin{tabular}{l|cccc|cccc}

\toprule
 {\bf Format} &  \multicolumn{4}{c}{{\bf W/o Tuning}} & \multicolumn{4}{c}{{\bf With Tuning}}  \\ 

  & Llama3.1 & Gemma2 & Qwen3 & Phi-4 & Llama3.1 & Gemma2 & Qwen3 & Phi-4 \\
  \midrule

  {\bf List} & {\bf 97.4} & 96.1 & 88.1 & {\bf 81.2} & 32.8 & 32.3 & 33.6 & {\bf 28.4} \\
  {\bf CSS} & 165.0 & 117.0 & 114.7 & 225.1 & 41.8 & 37.3 & 41.8 & 35.3 \\
  {\bf JSON} & 142.8 & {\bf 95.6} & {\bf 80.0} & 82.2 & 37.7 & {\bf 30.1} & 36.4 & 29.5 \\
  {\bf XML} & 121.4 & 101.4 & 104.5 & 85.3 & {\bf 31.8} & 31.7 & {\bf 32.8} & 29.7 \\

 \bottomrule
\end{tabular}
 \label{tab:results}
\end{table*}

Up until now, when prompting the LLM, we only include information about the landmark itself. However, it may be that a given landmark is situated close to several other landmarks. This can have the effect of altering the labeling instructions. This is intuitive, as it could be that the location where the label should be placed may be occupied by another landmark or another label itself. For example, in Figure~\ref{fig:label_placement_example} we can see that ``Azure Apartment'' is very close to ``Happy Trails''. This can affect the label placement of both landmarks so as they don't interfere with one another.

To account for the neighbors of a given landmark, we include them when: {\bf (1)} retrieving the labeling instructions from the database, {\bf (2)} prompting the LLM to place the label. When retrieving the instructions, this is simply done by including the names and types of all nearby landmarks. When prompting the LLM, the type and location of all nearby landmarks are further included in the prompt. The LLM is further instructed to consider these nearby landmarks when determining where to place the label. We detail the specific prompt in Figure~\ref{fig:neighbor_prompt}. In practice we only include those neighbors that are 50px or closer to the landmark being labeled.

\section{Experiments} \label{sec:exps}

In this section, we conduct extensive experiments to validate the effectiveness of the proposed method on the MAPLE dataset. Specifically, we attempt to determine: (RQ1) How well can various SOTA open-source LLMs perform? (RQ2) Does the performance vary by the type of format used to represent the coordinates and by the type of landmark? (RQ3) Can fine-tuning each LLM help improve performance? (RQ4) Can adding information about neighboring landmarks enhance the performance?   

\subsection{Experimental Settings} \label{sec:exp_setup}
{\bf Dataset}: The MAPLE dataset contains 1276 maps in total. They are split into training, validation, and test sets using an 80/10/20\% random split, respectively. The final number of landmarks for train, validation, and test is 883, 126, 267, respectively. Note that the training and validation maps are only used when instruction-tuning the LLMs.
\newline\newline
\noindent{\bf Models}: We test four different open-source and SOTA LLMs. This includes: Llama3.1 (8B)~\cite{grattafiori2024llama}, Gemma2 (9B)~\cite{team2024gemma}, Qwen3 (8B)~\cite{team2024gemma}, Phi-4 (14B)~\cite{abdin2024phi}.
\newline\newline
\noindent{\bf Tuning Strategy}: For each model, we also consider instruction tuning them to enhance their applicability towards the task of label placement. This is done through the use of QLoRA~\cite{dettmers2023qlora} for enhanced efficiency. We train for 5 epochs, with a learning rate of 1e-5 and a weight decay of 1e-5.
\newline\newline
\noindent{\bf Evaluation}: To evaluate the prediction quality, we consider the distance between the predicted location and the label. This is done by comparing the predicted location $\hat{y}_i$ and the centroid of the ground truth label $y_i$. Using these values for each landmark, we compute the RMSE across all samples $i=1$ to $N$.

\subsection{Main Results} \label{sec:exp_main}

\begin{figure}[h]
  \centering
  \includegraphics[width=0.6\linewidth]{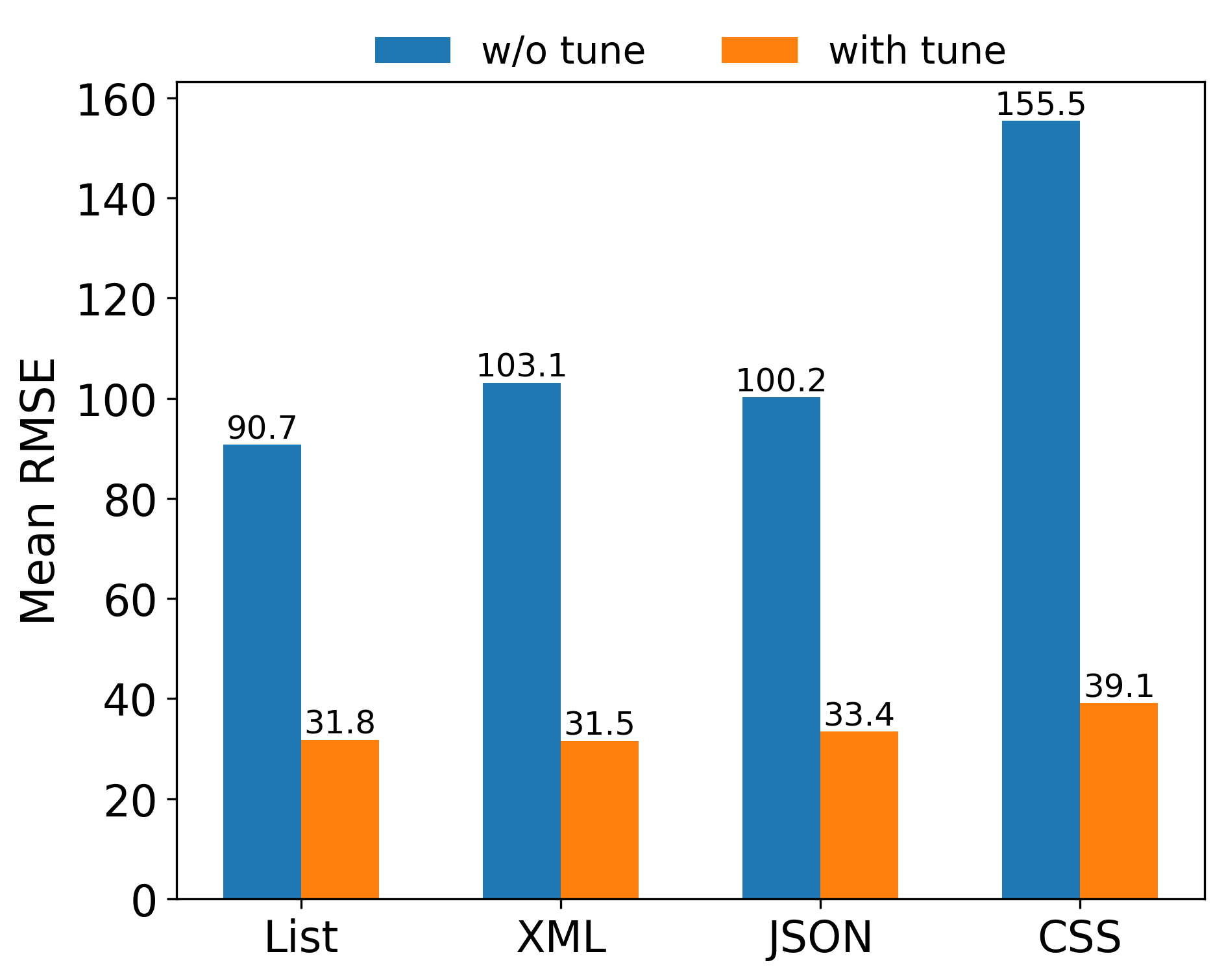}
  \caption{Mean performance across LLMs by type of coordinate format.}
  \label{fig:perf_by_format}
 \vskip -1.25em
\end{figure}

The main results are shown in Table~\ref{tab:landmark_results}. They are broken down by: LLM, with and w/o tuning, and by the location format. Inspecting the results, we can make a few observations. {\bf (1)} Some LLMs perform better than others. When not tuning, we can see that Qwen3 and Phi-4 tend to perform best, while with tuning, Phi-4 and Gemma 2 are best. {\bf (2)} Instruction tuning has a large positive effect on performance. Comparing the best performance of each LLM between the two strategies, their performance decreases by almost {\bf 200\%}. This indicates that the pre-trained LLM weights are likely not optimal for our task and require some additional supervision. {\bf (3)} We also observe that the location format has a very large effect on performance. In Figure~\ref{fig:perf_by_format}, we show the mean performance by format type across all LLMs. We can see that regardless of whether we tune the LLM, the performance varies considerably. Specifically, we find that the performance tends to be the most consistently strong when using the List format. On the other hand, formatting the coordinates via CSS tends to be the worst. Interestingly, however, when tuning, XML tends to perform best. These results underline that how we represent the location of each landmark does indeed matter and is an important consideration for ALP using LLMs. 

\subsection{Results by Landmark Type} \label{sec:perf_by_type}

We further display the results by type of landmark. As shown earlier in Table~\ref{tab:dataset_stats}, there are 7 different types of landmarks in our dataset. In order to understand the strengths and weaknesses of current LLMs for ALP, we must determine if it is better suited for labeling certain types of landmarks over others.

We show the results for the two best LLMs, Qwen3 and Phi-4 in Table~\ref{tab:landmark_results}. The results are displayed with and w/o tuning. We first observe that performance decreases across every type when tuning, suggesting that it benefits all types of landmarks. Second, both LLMs tend to struggle on both ``Leisure'' and ``Building'' landmarks. This is true even when tuning, despite the fact that ``Building'' landmarks have the second most number of samples in MAPLE. On the other hand, both LLMs tend to perform very well on those landmarks of type ``Shop'' and ``Office''. This suggests that the quality of ALP using LLMs can vary quite drastically by the type of landmark. Lastly, both LLMs are fairly consistent in their performance by type, with their overall trends being quite similar. 

\begin{table}[H]
\centering
 \caption{Results (RMSE) by type of landmark. Coordinates are formatted as a list.}
 \vskip -0.5em
\begin{tabular}{l|cc|cc}

\toprule
 {\bf Type} &  \multicolumn{2}{c}{{\bf Qwen3}} & \multicolumn{2}{c}{{\bf Phi-4}}  \\ 

  & w/o tune & with tune & w/o tune & with tune \\
  \midrule

    Tourism &        46.9	& 24.5	& 39.6	& 16.5 \\
    Shop & 26.7	& 13.9	& 27.7	& 8.7 \\
    Amenity &       60.4	& 22.9	& 61.1	& 23 \\
    Leisure & 137.7	& 52.4	& 127	& 36.3 \\
    Office & 21.9	& 8.9	& 23.5	& 5.7 \\
    Building & 116.2	& 41.6	& 111.2	& 39.9 \\
    Place &    41.4	& 22.3	& 40.3	& 10.3 \\
 \bottomrule
\end{tabular}
 \label{tab:landmark_results}
\end{table}

\subsection{Results with Neighboring Context} \label{sec:exp_neighbor}

We also experiment with adding the neighbors of a given landmark as described in Section~\ref{sec:method_neighbors}. In Table~\ref{tab:neighbor_results} we show the mean results with and w/o the neighboring context. We further display the results with and w/o tuning. For simplicity, we only test the List coordinate format, as it is the format that the LLMs perform most consistently well on.

Interestingly, we observe that there is often no benefit to this additional context. Specifically, outside of two scenarios (highlighted in {\bf bold}), the performance always slightly decreases. This raises the question of whether the neighboring landmarks are indeed helpful for labeling. It may also suggest that our current strategy for considering neighboring context is suboptimal and may need some refinement.

\begin{table}[H]
\centering
 \caption{Mean results (RMSE) across LLMs by w/o and with neighboring context. Coordinates are formatted as a list.}
 \vskip -0.5em
\begin{tabular}{l|cc|cc}

\toprule
 {\bf LLM} &  \multicolumn{2}{c}{{\bf W/o Tuning}} & \multicolumn{2}{c}{{\bf With Tuning}}  \\ 

  & std & + neighbors &  std & + neighbors \\
  \midrule

Llama3.1	& 97.4	& 104.5	& 32.8	& 36.9 \\
Gemma2	& 96.1	& 101.9	& 32.3	& {\bf 30.5} \\
Qwen3	& 88.1	& 90.3	& 33.6	& 35.7 \\
Phi-4	& 81.2	& {\bf 81.0}	& 28.4	& 29.5 \\
 \bottomrule
\end{tabular}
 \label{tab:neighbor_results}
\end{table}


\section{Conclusion}

In this paper we study the problem of automatic label placement (ALP) on maps. We find that no open-source datasets exist for benchmarking this task. To remedy this issue, we propose a new dataset, MAPLE, that contain 100 maps from three cities with over 1000 landmarks to be labeled. MAPLE also includes a set of labeling guidelines, meant to provide instructions on how to label various types of landmarks correctly. We experiment with using Large Language Models (LLMs) for this task where the relevant labeling instructions are retrieved using Retrieval Augmented Generation (RAG). We experiment with multiple prompting strategies along with instruction tuning. We show that LLMs can indeed perform ALP, however their performance differs by LLM and type of landmark. We further find that instruction tuning can help to dramatically improve the performance of LLMs for our task. For future work, we hope to experiment with other prompting strategies including in-context learning~\cite{incontextlearning}. We also hope to make use of the visual layout of the map itself through the use of vision language models (VLMs)~\cite{llava}.

\clearpage 
\bibliographystyle{ACM-Reference-Format}
\bibliography{ref}


\end{document}